\documentstyle[aps,prl,multicol]{revtex}

\begin {document}

\preprint{}
\title{Temperature-Dependent Pseudogaps in Colossal Magnetoresistive Oxides}
\author{T. Saitoh,$^1$\cite{adrT} D. S. Dessau,$^1$ Y. 
Moritomo,$^2$\cite{adr} T. Kimura,$^{2}$ Y. Tokura$^{2,3}$ and N.
Hamada$^{2}$\cite{adrH}}
\address{$^1$Department of Physics, University of Colorado, Boulder, Colorado
80309-0390}
\address{$^2$Joint Research Center for Atom Technology, Tsukuba, Ibaraki 
305-0046,
Japan}
\address{$^3$Department of Applied Physics, University of Tokyo, Tokyo 
113-0033,
Japan}
\date{Received \today}
\maketitle
\vspace*{0.1 in}
%--------1-------------------3-------------------5-------------------7--------

\begin{abstract}
\vspace*{-0.3 in}

Direct electronic structure measurements of a variety of the colossal
magnetoresistive oxides show the presence of a pseudogap at the Fermi
energy $E_F$
which drastically suppresses the electron spectral function at $E_F$.
The pseudogap is a strong function of the layer number of the samples
(sample dimensionality)
and is strongly temperature dependent, with the changes beginning at the
ferromagnetic
transition temperature $T_c$. These trends are consistent with the major
transport trends
of the CMR oxides, implying a direct
relationship between the pseudogap and transport,
including the  ``colossal" conductivity changes which occur
across $T_c$. The $k$-dependence of the temperature-dependent effects 
indicate that the pseudogap observed in these compounds is not due to 
the extrinsic effects proposed by Joynt. 

\end{abstract}

%79.60.-i: Photoemission and photoelectron spectra
%71.30.+h: Metal-insulator transitions and other electronic transitions
\vspace*{-0.3 in}

\begin{multicols}{2}
\narrowtext

The colossal magnetoresistance (CMR) effect recently discovered in the
manganese oxides (La$_{1-x}B_x$MnO$_3$ ($B$=Sr, Ca, Ba) and
La$_{2-2x}$Sr$_{1+2x}$Mn$_2$O$_7$) is a phenomenon which dramatically displays
the  strong correlation between magnetism and electronic
conduction\cite{CMR,Moritomo}. At low temperatures ($T$), properly doped 
manganese
oxides exhibit ferromagnetic metallic behavior, while at high $T$ they
exhibit paramagnetic insulating behavior. If one starts at $T$
slightly above $T_c$, the application of a magnetic
field can play a similar role as $T$ by driving the material through
the insulator-metal transition.  This is the CMR effect.

Traditionally, the starting point for understanding the
electronic and magnetic properties of the CMR oxides is the double exchange
(DE) model originally studied by Zener, de Gennes,
and Anderson and Hasegawa\cite{ZenerETAL}. 
DE says that the hopping probability $t$ for an $e_g$ 
symmetry (conduction)
electron to hop from one site to the next is $t=t_0\cos(\theta/2)$, 
where $t_{0}$ is the bare hopping probability and $\theta$ is the 
relative angle between two core ($t_{2g}$) spins (see Fig.~\ref{fig1}).
A ferromagnetically ordered sample has $\theta$=0 and so a full hopping
probability, while an antiferromagnetically ordered sample has
$\theta=180^\circ$ and no probability of hopping. The paramagnetic case
corresponding to $T>T_c$ can be approximated by $\theta=90^\circ$, i.e.
$t$ should be reduced to $\cos(90^\circ/2)$ or 
about 70\% of its original value.  A calculation of the bandwidth 
($W$) change with
temperature is done more precisely by Kubo \cite{Kubo}, with the result
coming out very similarly to our simple explanation. Directly associated
with the change in hoppping probablity is the effect on the
$E$ vs. $k$ relations and $N(E_F)$, which is schematically illustrated in
Fig.~\ref{fig1}(B) and (C)\cite{Kubo}.  
The roughly 30\% changes in $t$ and $W$ expected
across the transition will contribute to similar changes in the electronic
mobility $\mu$ and conductivity $\sigma$.  Since the conductivity
changes across the transition may be many orders of magnitude instead of
just a 30\% effect (see Fig.~\ref{fig6}(B) and reference\cite{Moritomo}), 
it has been
argued that additional physics must be necessary to explain the conductivity
changes in the manganites\cite{Millis,Roder}. The results presented here
confirm these ideas but go one farther, indicating that at least for 
the layered samples, DE is probably not
even the dominant mechanism but is instead supplementing some other more 
important physics.  

We performed Angle Resolved Photoemission (ARPES) experiments on 
cleaved single crystalline samples of the layered and pseudocubic 
manganites at the Stanford Synchrotron
Radiation Laboratory (SSRL) using a 50 mm hemispherical analyzer. The
energy resolution was about 40 meV FWHM and the $k$ resolution was better
than $\pm$0.05$\pi$ in the 1st Brillouin zone at the photon energy
of 22.4 eV.
The samples were grown by the floating-zone
method\cite{Moritomo,Urushi}.
The surfaces of the layered samples were
mirror-like and high quality low energy electron diffraction patterns
were easily obtainable, with no evidence of extra superlattice spots.
Other indications
of the quality of the surfaces are the large amount 
of $E$ vs. $k$ dispersion
observed and the $T$ dependence, which as we will see has
many similarities to the Drude part of optical conductivity mesurements
whose probing
depth is typically several thousands of \AA \cite{Okimoto}. 
We note that the recent claim that
the surface magnetism of the manganites is different from that of the
bulk\cite{JHPsurface} is on
strained thin films which have undergone a complicated surface
cleaning procedure.  We consider these films to have a non-ideal
surface, which is further confirmed by a lack of $E$
vs. $k$ dispersion from these surfaces.

Figure~\ref{fig2} shows the full valence band spectra of
La$_{1.2}$Sr$_{1.8}$Mn$_2$O$_7$ along the
(0,0)$-$($\pi$,0)$-$($\pi$,$\pi$) symmetry line at 200 K (paramagnetic
state) and at 50 K (ferromagnetic state).  Figure~\ref{fig3}(A) shows 
the near-$E_{F}$ data from the same sample taken under identical 
conditions.  All the spectra have been normalized by the total area of the main valence-band spectra
at the highest $T$ at each $k$-space point. The procedure for the
angle-integrated spectra of cubic compounds is the same.
All 200K data were taken first, immediately after the 
sample cleave which was also performed at 200K.  At each emission 
angle the valence band and near-$E_F$ spectra were taken concurrently 
without a sample reoptimization or realignment.  After all 200K data 
were finished the sample was cooled to 50K and the measurement 
procedure was repeated.  Therefore the low temperature spectra are 
more aged than the high temperature spectra, accounting for the slightly 
higher (but still very small)
``dirt'' peak at -10 eV observed in the low temperature spectra.  We note 
that with continued ageing the dirt peak continues to grow and the 
emission intensity near $E_{F}$ is found to decrease.  The fact that 
our low temperature spectra show the same or even more weight than our 
high temperature spectra implies that the temperature dependent 
reduction in the spectral weight at high temperatures discussed here 
is not due to an ageing effect.  We also
checked the reproducibility of the $T$-dependent changes by several
cleaves including warming runs and have confirmed that the qualitative
changes were much larger than the aging effects.

Figure~\ref{fig3}(A) shows two major features, labeled $\alpha$ and
$\beta$, the centroids of which are plotted as a function of $k$ in
Fig.~\ref{fig2}(B) which also includes the $e_{g\uparrow}$ bands from 
our LSDA+$U$
band structure calculations \cite{Dessau}. According to the calculation as
well as our polarization-depedendent photoemission experiments 
\cite{DessauMRS},
$\alpha$ has primarily $d_{x^2-y^2\uparrow}$ symmetry and $\beta$ has
primarily $d_{3z^2-r^2\uparrow}$ symmetry. We see that the dispersion and
energy position of the centroids of the ARPES features is in relatively
good agreement with the calculations except near
$E_F$, where a pseudogap suppresses the weight at $E_{F}$. Additionally, 
the experiment shows a k-space locus of lowest energy 
excitations forming a large "ghost" Fermi surface (FS),\cite{Dessau}
which is in rough agreement with FS calculated by the LSDA band 
theory. This large hole FS is also consistent with 
recent Hall effect measurements.\cite{Hall}

Robert Joynt has recently argued that the
pseudogap observed in photoemission experiments in the manganites and
possibly many other compounds may be due to extrinsic loss
effects\cite{Joynt}.  It is important that we can exclude this 
possibility before analyzing our data in more detail.  In Joynt's 
proposal, the electric field created by the
outgoing photoelectron produces ohmic losses, lowering 
the kinetic energy of the ejected photoelectron.  This would show up 
in the spectra as a pseudogap,
even if there was not one in the true density of states, and should 
be most important for highly resistive samples.  

The data of figure 3(A) can directly address this question. Some of the 
curves (for example $\vec k = (\pi,.28\pi)$)
show a significant shift away from $E_{F}$
with temperature, while other curves (for example $\vec
k =(.46\pi,0)$) show a minimal effect.
The minimal temperature-dependent changes observed at $(.46\pi,0)$
put an upper limit on the magnitude of the effects that could be due
to Joynt's extrinsic losses, and so the dramatic changes observed at $\vec
k = (\pi,.28\pi)$ should be of intrinsic origin.  Additionally, since 
there is no effect at some angles from a two-order of magnitude change of
resistivity, it should also be
clear that the weaker pseudogap observed at $(\pi,.28\pi)$ at 50K in 
the spectra is not due to ohmic losses (the pseudogap should be
measured at the $k$ where 
the peak is closest to $E_{F}$, which occurs at $(\pi,.28\pi)$).  In a 
recent comment and reply, \cite{DanComm,Reply} Joynt 
agrees that the spectral weight suppression in these compounds appears 
to be intrinsic.

We discuss two main aspects of the $T$-dependence of the
data of Fig.~\ref{fig2} and 3: (1) The data shows only a very small change 
in $W$, in contradiction with the DE
prediction. (2) There are large changes in the spectral intensity at
$E_F$ for $k$ near ($\pi$,0.28$\pi$), which is due to the opening of a
pseudogap. 

(1) DE tells us that $W$ should decrease by
about 30\% when going from the ferromagnetic to the paramagnetic 
phases (see Fig.~\ref{fig1}).
In our measurement, the highest binding energy
occupied states that are clearly resolvable are at (0.18$\pi$,0) and a
binding energy near 1.6 eV.  Upon going to the high $T$
paramagnetic state, DE theory predicts that these states should show a
significant ($\sim$30\%) energy shift towards $E_F$, i.e. they should
appear centered around 1.1 eV. Instead, we observe a small energy shift
of approximately 0.06 eV, or a change in $W$ of just a few
percent (0.06/1.5=4\%). This lack of a change in $W$
should imply that a short-range (in-plane) ferromagnetic correlation
still exists above $T_c$.
This is consistent with a recent neutron scattering
measurement from the same sample source which showed in-plane short-range
ferromagnetic order as high as $T$=284 K \cite{Perring}.

(2) In contrast to the small $T$ dependent changes at (0.18$\pi$,0), there 
are dramatic $T$ dependences near
$k$= ($\pi$,0.28$\pi$), which corresponds to a predicted FS crossing point
as well as the point of closest approach of the peak centroid to $E_F$.
The spectra at the higher $T$ is pushed farther away
from $E_F$ and the spectral intensity near $E_F$ is reduced. This behavior
represents the opening of a (pseudo)gap centered at $E_F$ with $T$. 
In particular, a gap centered at $E_F$ is expected to affect the states near
$k_F$ strongly and those away from $k_F$ much more weakly, as observed in
our data. We call this gap a pseudogap because the edges of the gap are
``soft" and because the spectral weight at $E_F$ is not always completely
suppressed.

We note that even at low $T$ the spectra at ($\pi$,0.28$\pi$)
remain pulled back from $E_F$, and the spectral weight reaching $E_F$
is vanishingly small. This indicates that the pseudogap is active at
low as well as at high $T$, although it is stronger at high
$T$. Figure~\ref{fig4}(A) shows this $T$ dependence in more
detail from a different La$_{1.2}$Sr$_{1.8}$Mn$_2$O$_7$ sample, also 
measured while cooling. The triangles indicate the spectral weight found very near
$E_F$ obtained over two different energy integration windows, and the
diamonds show the energy shift of the leading edge (at two different
positions on the edge) as a function of $T$.  It is seen that
both the near-$E_F$ weight and the position of the leading edge begin
increasing at $T_c$ and then rise monotonically without saturation as
the $T$ is lowered.

Figure~\ref{fig4}(B) shows that a similar $T$-dependent behavior is
found for the perovskite samples, although this time the $T_c$ is
very different and we plot the angle-integrated DOS\cite{3dDOS}.  As 
in figures 2 and 3, the high temperature data was measured prior to 
the low temperature data, negating the possibility that the trends 
observed here are due to ageing.
A portion of the raw data is
shown in the upper right hand corner of Fig.~\ref{fig3}(A), with similar raw
data being presented by other groups\cite{JHP,Sarma}. The high
$T$ data does not exhibit any weight right at $E_F$, while
the low $T$ data exhibits a clear Fermi-edge cutoff indicating
a finite DOS at $E_F$.
 
The left vertical scale of Fig.~\ref{fig4}(B)
corresponds to the ratio of the observed weight to the weight deduced
from the band structure calculations\cite{Pickett}.  The way in which 
this ratio was determined is shown more clearly in figure 5, which 
shows a comparison of low temperature (22K) experimental angle-integrated valence band data to 
a photoionization cross-section-weighted band 
structure density of states\cite{Yeh}.  To make this comparison, a small ($\sim$15\%) 
background due to inelastically
scattered electrons was removed from the experimental 
photoemission data, and the integrated spectral area under the two 
curves from 0 to 8 eV was made to match.

While the agreement between the theory and experiment is nowhere very 
good, it is worst near $E_{F}$, where the experiment shows a 
greatly reduced weight compared to theory.  The left axis of figure 4 
shows that the near-$E_F$ weight always remains at least a factor of 10
lower than the predicted weight, even at the lowest $T$, even
though a clear Fermi edge is observed in the perovskite spectrum of
Fig.~\ref{fig3}(A). As shown in the figure, a similar $T$ dependence
and reduction in low frequency spectral weight is observed in the
Drude portion of optical conductivity experiments of a very similar
($x$=0.175) sample\cite{Okimoto}. 

Figure~\ref{fig6} shows near-$E_F$ photoemission data and DC resistivity from
the three different families of the manganites, all with the same
doping level of 0.4 holes per Mn site. At the low 
$T$ of the measurements, the cubic sample has a finite $N(E_F)$ and is
metallic, the $n$=1 sample has a clear gap and is
insulating ($n$=1 samples are insulating for all doping levels and
$T$\cite{MoritomoPRB}), while the $n$=2 sample has a
vanishingly small weight at $E_F$ and is barely metallic. The trends
in the data again indicate the significance of the pseudogap in
explaining the transport properties of the manganites.
Our data indicates a direct and possibly causal 
relationship between
$N(E_F)$ and conductivity $\sigma$.  Considering the very small
$W$ changes that are observed across $T_c$, it appears
that the changes in $N(E_F)$ due to the pseudogap are more important
for the metal-insulator transition than is the DE effect.

It may be tempting to attribute the decrease in spectral weight at $E_F$
to be due to a decrease in mobile carriers. For instance, on the basis
of transport measurements Jaime {\it et al.} have proposed a two-fluid
model where the reduced conductivity of the manganites is due to a
$T$-dependent reduction in the number of otherwise normal high
mobility carriers  \cite {Jaime}.  Such a situation should lead to a
small FS with normal (i.e. non-gapped) behavior. Here, we have a large
``ghost" FS enclosing a large number of carriers\cite{Dessau}, although each of these
carriers gives a drastically reduced spectral weight at $E_F$ due to
the pseudogap. 
The large size of this observed ghost FS
is consistent with the recent Hall effect experiments which indicated
a large FS.\cite{Hall}

A few theoretical works are able to predict gaps or
pseudogaps in the electron spectral function.  In particular,
the Jahn-Teller (J-T) effect has been heavily discussed in the context of
the manganites.  Strong coupling between the electrons and the J-T
phonons will break the degeneracy of the $e_{g}$
orbitals, opening a gap between them.\cite{Millis,Singh}
However, $E_{F}$ will only lie in this gap
for a doping level of $x=0$, i.e. for the $d^{4}$ compounds such as
LaMnO$_{3}$. Our work clearly shows that there is a gap
at $E_{F}$ for filling levels far away from $x=0$, so coupling to Jahn-Teller
phonons can not explain the pseudogap.  The inclusion of other
phonons such as breathing-mode phonons may be relevant,\cite{Millis} although it
is unclear if the electron-phonon coupling to these modes can be
strong enough to open a gap as large as what we have experimentally
observed.  If so, then the pseudogap as well as the broad ARPES peaks
may be explainable within the context of strong electron-phonon
coupling\cite{Dessau}. Other interesting possibilities are the recent works by
Alexandrov et al\cite{Alexandrov} and Moreo et al\cite{Moreo}, both of
which predict the gap to exist at $E_{F}$
irrespective of the doping level, as observed experimentally.
Alexandrov's pseudogap is due to the
formation of bipolarons, while Moreo's is due to strong
electron-phonon coupling in the presence of phase seperation.  
Other effects that should be considered are the charge/orbital ordering
tendencies\cite{order} as well as long-range Coulomb interactions
which may open a Coulomb gap\cite{Varma}.

Finally, we briefly contrast the physics between the pseudogap discussed here 
and the pseudgap in the high temperature superconductors (HTSC's) 
\cite{HTC pseudogap}. In the HTSC's the pseudogap occurs below the 
temperature T* and corresponds to a reduced resistance state, while in 
the manganites the pseudogap is most active above the temperature $T_{c}$, 
and corresponds to 
an increased resistance state.  We suggest that this can be 
understood by assuming that in the HTSC's only the spin excitations are gapped, while for the 
manganites the charge excitations are gapped as well.  The elucidation of the pseudogap
origin will likely be a crucial step in our
progress to understand the physics of both of these
families of compounds.

We would like to thank A Andreev, A. Fujimori, T.
Katsufuji, A. Millis, T. Mizokawa, C.-H. Park, L. Radzihovsky and Z.-X. Shen for fruitful discussions
and W. E. Pickett for the numerical data of the band-structure
calculations. SSRL is operated by the DOE, Office of Basic Energy Sciences. 
Other support came from an ONR Young Investigator grant and from NEDO.

%Fig.1
\begin{figure}
  \caption{(A) The relevant Mn-O derived electronic orbitals for the
manganites. According to DE 
theory $t$ from one site to the next is a strong
function of the relative angle ($\theta$) between the $t_(2g)$ spins on the
two sites. (B) Schematic drawing of the 
$E$ vs. $k$ relations expected within DE theory. The paramagnetic states 
should have a dispersion about 70\% of that in the ferromagnetic state. (C) 
Total electronic density of states $N(E)$ expected within DE theory.  
Concomitant with the reduced dispersion and bandwidth, the paramagnetic 
phase is expected to have a slightly higher $N(E)$ at $E_F$.}
\label{fig1}
\end{figure}

%fig.2
\begin{figure}
  \caption{(A) ARPES spectra from $h\nu$=22.4 eV of the full valence band region from
La$_{1.2}$Sr$_{1.8}$Mn$_2$O$_7$ along the
(0,0)$-$($\pi$,0) (left panel) and ($\pi$,0)$-$($\pi$,$\pi$) (right panel)
symmetry lines at 200K (red) and 50K (blue).
The square panels at top show the location of the cuts in one quadrant 
of the two-dimensional
Brillouin zone.}
  \label{fig2}
\end{figure}

%fig.3
\begin{figure}
  \caption{(A) Near-$E_{F}$ spectra taken under identical conditions 
  as the spectra of figure 2. 
  The red $\times$ shows the predicted FS crossing point.
In the right panel, the La$_{0.82}$Sr$_{0.18}$MnO$_3$ spectra at two
$T$ and the gold spectrum are shown. (B) Comparison between the
experimental dispersion of La$_{1.2}$Sr$_{1.8}$Mn$_2$O$_7$ from Panel A
(red: 200 K, blue: 50 K) and the LSDA+$U$
bands (green).}
  \label{fig3}
\end{figure}

%Fig.4
\begin{figure}
  \caption{(A) Integrated spectral weight of La$_{1.2}$Sr$_{1.8}$Mn$_2$O$_7$
in two different windows and relative energy shifts of the peak and the edge
at -0.3 eV at ($\pi$,0.28$\pi$). (B) Comparison of the integrated spectral
weight of La$_{0.82}$Sr$_{0.18}$MnO$_3$ in two different windows between the
experimental spectra and the band-structure
calculations\protect\cite{Pickett}.
The Drude weight of $x$=0.175 sample\protect\cite{Okimoto} is also shown for
comparison.}
  \label{fig4}
\end{figure}

%Fig.5
\begin{figure}
  \caption{Angle integrated valence band data from  
  La$_{.82}$Sr$_{.18}$MnO$_3$ (red) versus the cross-section corrected 
  band structure density of states, courtesy of W. Pickett.}
  \label{fig5}
\end{figure}

%Fig.6
\begin{figure}
  \caption{(A) Normal emission photoemission data in the near $E_F$ region
for the layered ($n$=1, 2) and cubic ($n=\infty$) systems, all with doping
$x$=0.4. 
%The zeroes have been vertically offset for clarity. 
For the layered
samples, this $k$-point mostly samples the $d_{3z^2-r^2}$ out-of-plane bands
($\beta$ in Fig.~\protect\ref{fig2}). The photon energy was 48 eV for $n$=1,2 
and 36 eV for $n=\infty$ and for Au. At these high photon energies these 
states 
have a higher cross section than shown in Fig.~\protect\ref{fig2}.  
Spectra taken at $k$ near ($\pi,0.28\pi$) show a similar layer-number 
dependence (not shown). (B) Resistivity as a function of $T$ for the same 
samples\protect\cite{Moritomo}.}

      % n=1  MN75-006f   0/0  hv=48eV  Ep=2 60x30  T=22K
      % n=2  Mn83-038f   0/0  hv=48eV  Ep=5 60x30  T=45K
      %n=infinity  Mn70-026f  0/0  hv=36eV  Ep     T=21K
      %Au - Mn70-027Auf  Ef=32.069  deltaE=69meV   T=21K
   \label{fig6}
 \end{figure}

\end{multicols}


\begin{references}
\bibitem[*]{adrT}Present address: Photon Factory, KEK, Tsukuba 305-0801, Japan.
\bibitem[**]{adr}Present address: CIRSE, Nagoya University, Nagoya 464-01,
Japan.
\bibitem[***]{adrH}Present address: Science University of Tokyo,
Noda 278-8510, Japan
%1
\bibitem{CMR}K. Chahara {\it et al.}, Appl. Phys. Lett. {\bf 63},
1990 (1993); R. von Helmolt {\it et al.}, Phys. Rev. Lett. {\bf 71}, 2331
(1993); Y. Tokura {\it et al.}, J. Phys. Soc. Jpn. {\bf 63}, 3931 (1994).
%2
\bibitem{Moritomo}Y. Moritomo {\it et al.}, Nature {\bf 380}, 141 (1996).
%3
\bibitem{ZenerETAL}C. Zener, Phys. Rev. {\bf 82}, 403 (1951), P.-G. 
de Gennes, Phys. Rev. {\bf 118}, 141 (1960), P. W. Anderson and 
H. Hasegawa, Phys. Rev. {\bf 100}, 675 (1955).
%4
\bibitem{Kubo} K. Kubo, J. Phys. Soc. Jpn. {\bf 33}, 929 (1972).
%5
\bibitem{Millis} A.J. Millis, P.B. Littlewood, B.I. Shraiman, Phys. Rev. Lett.
{\bf 74}, 5144 (1995); A.J. Millis, B.I. Shraiman, R. Mueller, {\it ibid.},
{\bf 77}, 175 (1996); A.J. Millis, R. Mueller, B.I. Shraiman, Phys.
Rev. B {\bf 54}, 5405 (1996).
%6
\bibitem{Roder} H. R\"oder,J. Zang,and A. Bishop, Phys. Rev. Lett.
{\bf 76}, 1356 (1996).
%7
\bibitem{Urushi}A. Urushibara {\it et al.}, Phys. Rev. B {\bf 51}, 14103
(1995).
%9
\bibitem{Okimoto}Y. Okimoto {\it et al.}, Phys. Rev. B {\bf 55}, 4206 (1997).
%10
\bibitem{JHPsurface}J.-H. Park {\it et al.}, Phys. Rev. Lett. {\bf 81},
1953 (1998)
%11
\bibitem{Dessau} D.S. Dessau {\it et al.}, Phys. Rev. Lett. {\bf 81}, 192
(1998).
%12
\bibitem{DessauMRS} D.S. Dessau {\it et al.}, Science and Technology of 
Magnetic Oxides, Materials Research Society Symposium Proceedings {\bf 
494},181 (1998)
\bibitem{Hall} A. Asamitsu and Y. Tokura , Phys. Rev. B {\bf 58}, 47
(1998).
\bibitem{Joynt}R. Joynt, Science {\bf 284}, 777 (1999).
\bibitem{DanComm}D. S. Dessau and T. Saitoh, Science (to appear).
\bibitem{Reply}R. Joynt, Science (to appear).

%16
\bibitem{Perring} T. Perring {\it et al.}, Phys. Rev. Lett. {\bf
80}, 4359 (1998)
%17
\bibitem{3dDOS} We have not been able to obtain the mirror-quality cleaves 
necessary to obtain high quality ARPES data on the perovskite samples.
%18
 \bibitem{JHP}J.-H. Park {\it et al.}, Phys. Rev. Lett. {\bf 76}, 4215
(1996). {\it ibid.} {\bf 81}, 1953 (1998), Nature (London) {\bf 392},
 794 (1998).
%19
\bibitem{Sarma}D. D. Sarma {\it et al.}, Phys. Rev. B {\bf 53}, 6874 (1996).
%21
\bibitem{Pickett} W. E. Pickett and D. J. Singh, Phys. Rev. B {\bf 53},
1146 (1996).

\bibitem{Yeh} The Mn 3d and O 2p partial density of states from the 
band theory calculation were each individually weighted by the 
theoretical cross sections, obtained from J. J. Yeh and I. Lindau, 
At. Data Nucl. Data Tables {\bf 32}, 1 (1985).  Following this a 
slight energy-dependent broadening was applied to the theoretical 
density of states to simulate lifetime and energy resolution effects.

\bibitem{MoritomoPRB} Y. Moritomo {\it et al.}, Phys. Rev. B {\bf 51}, 3297
(1995).
%23
\bibitem{Jaime}  M. Jaime and M. Salamon, ``A two-Fluid Model for
Manganites'',Telluride Summer Workshop on Magnetoresistive Oxides, July 1998.

%24
\bibitem{Singh} D.J. Singh {\it et al.}, Phys. Rev. B {\bf 57}, 88 (1998).
%25
\bibitem{Alexandrov} A.S. Alexandrov and A.M. Bratkovsky, Phys. Rev.
Lett {\bf 82}, 141 (1999).
%26
\bibitem{Moreo} A. Moreo {\it et al.}, Phys. Rev. Lett. {\bf
83}, 2773 (1999).
%27
\bibitem{order} S. Ishihara{\it et al.}, Phys. Rev. B {\bf 55}, 8280 (1997).
%28
\bibitem{Varma} C.M. Varma, Phys. Rev. B {\bf 54}, 7328 (1996).
%29
%30
\bibitem {HTC pseudogap} B.G. Levi, Physics Today, p.17 June 1996.

\end{references}
\end{document}